%% file: MIMO_dfe_wcnc_arxiv.tex
\documentclass[10pt,twocolumn,conference]{IEEEtran}
\usepackage{times}
\usepackage{amsbsy}
\usepackage{latexsym}
\usepackage{amsmath}
\usepackage[ansinew]{inputenc}
\usepackage[dvips ]{graphicx}
\input epsf
\usepackage{epic,eepic}
\newcommand{\ms}{\;\;}

\title{\huge Adaptive Decision Feedback Reduced-Rank Equalization Based on
Joint Iterative Optimization of Adaptive Estimation Algorithms for
Multi-Antenna Systems}
\author{Rodrigo C.\ de Lamare \dag, Are Hj{\o}rungnes \ddag ~ and Raimundo Sampaio-Neto \S \\
\fontsize{10}{10}\selectfont\itshape \dag Communication Research Group, Department of Electronics, The University of York, UK \\
 \ddag UniK - University Graduate Center,
University of Oslo, Norway \\ \S CETUC/Pontifical Catholic University of Rio de Janeiro (PUC-RIO), Brazil \\
\fontsize{9}{9}\selectfont\ttfamily\upshape E-mails:
rcdl500@ohm.york.ac.uk, arehj@unik.no, raimundo@cetuc.puc-rio.br
\thanks{\footnotesize This work was
supported by the Research Council of Norway VERDIKT project
176773/S10 called OptiMO.}
}

\begin{document}

\maketitle

\begin{abstract}
This paper presents a novel adaptive reduced-rank
multi-input-multi-output (MIMO) decision feedback equalization
structure based on joint iterative optimization of adaptive
estimators. The novel reduced-rank equalization structure consists
of a joint iterative optimization of two equalization stages,
namely, a projection matrix that performs dimensionality reduction
and a reduced-rank estimator that retrieves the desired
transmitted symbol. The proposed reduced-rank structure is
followed by a decision feedback scheme that is responsible for
cancelling the inter-antenna interference caused by the associated
data streams. We describe least squares (LS) expressions for the
design of the projection matrix and the reduced-rank estimator
along with computationally efficient recursive least squares (RLS)
adaptive estimation algorithms. Simulations for a MIMO
equalization application show that the proposed scheme outperforms
the state-of-the-art reduced-rank and the conventional estimation
algorithms at about the same complexity.\\


\end{abstract}

\section{Introduction}

\PARstart{T}{he} high demand for performance and capacity in
wireless networks has led to the development of numerous signal
processing and communications techniques for employing the
resources efficiently. Recent results on information theory have
shown that it is possible to achieve high spectral
efficiency~\cite{foschini} and to make wireless links more
reliable \cite{alamouti,tarokh1} through the deployment of
multiple antennas at both transmitter and receiver. In MIMO
communications systems, the received signal is composed by the sum
of several transmitted signals which share the propagation
environment and are subject to multiple propagation paths and
noise at the receiver. The multipath channel originates
intersymbol interference (ISI), whereas the non-orthogonality
among the signals transmitted gives rise to multi-access
interference (MAI) at the receiver.

In order to mitigate the detrimental effects of ISI and MAI, that
reduce the performance and the capacity of MIMO systems, the
designer has to construct a space-time and MIMO equalizer. The
optimal MIMO equalizer known as the maximum likelihood sequence
estimation (MLSE) receiver was originally developed in the context
of multiuser detection in \cite{verdu}. However, the exponential
complexity of the optimal MIMO equalizer makes its implementation
costly for multipath channels with severe ISI and MIMO systems
with many antennas. In practice, designers often prefer the
deployment of low-complexity MIMO receivers such as the linear
\cite{duel,tehrani} and MIMO decision feedback equalizers (DFE)
\cite{dhahir,kominakis}. Among these, the DFE
\cite{dhahir,kominakis} can offer substantially better performance
than their linear counterparts due to the interference
cancellation capabilities of the feedback section for good channel
conditions. These receivers require the estimation of the
coefficients used for combining the received data and extracting
the desired transmitted symbols. A challenging problem which
remains unsolved by conventional estimation techniques is that
when the number of elements in the estimator is large, the
algorithm requires substantial training for the MIMO DFE and a
large number of received symbols to reach its steady-state
behavior.

Reduced-rank estimation \cite{scharf}-\cite{delamaresp} is a very
powerful and effective technique in low-sample-support situations
and in problems with high-order estimators.  The advantages of
reduced-rank estimators are their faster convergence speed and
better tracking performance than existing techniques when dealing
with large number of weights. Several reduced-rank methods and
systems have been proposed in the last several years, namely,
eigen-decomposition techniques \cite{song&roy}, the multistage
Wiener filter (MWF) \cite{goldstein}, and the auxiliary vector
filtering (AVF) algorithm \cite{avf}. Prior work on reduced-rank
estimators for MIMO systems is extremely limited and relatively
unexplored, being the work of \textit{Sun et al.} \cite{sun} one
of the few existing ones in the area.

In this work, we propose a novel adaptive MIMO decision feedback
equalization structure based on a novel reduced-rank estimation
method. The proposed reduced-rank equalization structure consists
of a joint iterative optimization of two equalization stages,
namely, a projection matrix that performs dimensionality reduction
and a reduced-rank estimator that retrieves the desired
transmitted symbol and is then followed by a feedback section that
is responsible for cancelling the multi-access interference caused
by the associated users. The essence of the proposed scheme is to
change the role of the equalization filters of the feedforward
section of the MIMO DFE. The projection matrix is responsible for
performing dimensionality reduction, whereas the reduced-rank
estimator effectively retrieves the desired signal. In order to
estimate the coefficients of the proposed MIMO reduced-rank DFE,
we describe least squares (LS) expressions for the design of the
projection matrix and the reduced-rank estimator along with
computationally efficient recursive least squares (RLS) adaptive
estimation algorithms. The performance of the proposed adaptive
MIMO reduced-rank DFE and estimation algorithm is assessed and
compared with the best known estimation schemes via numerical
simulations.

The rest of this paper is structured as follows: The MIMO system
model is given in Section II. The conventional adaptive MIMO DFE
is reviewed in Section III, whereas the proposed adaptive MIMO
reduced-rank DFE is introduced in Section IV. Section V is devoted
to the development of the LS estimators and the computationally
efficient RLS algorithms. Section VI presents and discusses the
simulation results and Section VII gives the concluding remarks of
this work.

\section{MIMO System Model}

\begin{figure}[h!]
  \begin{center}
 \def\eepicsize#1#2{1\columnwidth}
 \input{./fig1.eepic}
    \caption{MIMO system model.}
    \label{fig:system}
  \end{center}
\end{figure}
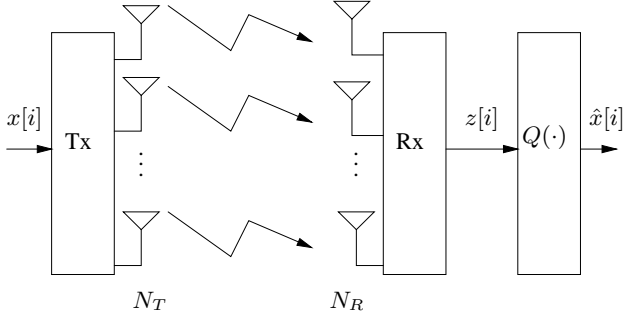

Consider a MIMO system with $N_T$ antennas at the transmitter and
$N_R$ antennas at the receiver in a spatial multiplexing
configuration, as shown in Fig.~\ref{fig:system}. The signals are
transmitted from $N_T$ antennas over multipath channels with $L_p$
propagation paths and are received by $N_R$ antennas. We assume
that the channel is constant during each packet transmission and
the receiver is synchronized with the main path.

The received signals are filtered by a matched filter, sampled at
symbol rate, organized in a window of $L$ symbols ($L > L_p$) for
each antenna element and yield the $L N_R \times 1$ received
vector
\begin{equation}
\label{eq:one} {\bf y}[i] = {\bf H}[i] {\bf x}_T[i] + {\bf n}[i],
\end{equation}
where ${\bf y}[i] = \big[ {\bf y}_1^T[i] ~ {\bf y}_2^T[i] ~ \ldots
~{\bf y}_{N_R}^T[i] \big]^T$ contains the signals collected by the
$N_R$ antennas, the $L \times 1$ vector ${\bf y}_k[i] = \big[
y_{k,1}[i] ~ y_{k,2}[i] ~ \ldots ~ y_{k,L}[i] \big]^T$, for $k=1,
~\ldots,~ N_R$, contains the signals collected by the $k$th
antenna and are organized into a vector. The $L N_R \times L N_T$
MIMO channel matrix ${\bf H}[i]$ is described by
\begin{equation}
\label{eq:two} {\bf H}[i] = \left[ \begin{array}{cccc}
  {\bf H}_{1,1}[i] & {\bf H}_{1,2}[i] & \ldots & {\bf H}_{1,N_T}[i] \\
  {\bf H}_{2,1}[i] & {\bf H}_{2,2}[i] & \ldots & {\bf H}_{2,N_T}[i] \\
  \vdots & \vdots & \ddots & \vdots \\
  {\bf H}_{N_R,1}[i] & {\bf H}_{N_R,2}[i] & \ldots & {\bf H}_{N_R,N_T}[i]
\end{array}\right],
\end{equation}
where the $L \times L$ matrix ${\bf H}_{k,j}[i]$ describes the
multipath channel from antenna $k$ to antenna $j$. The $L N_T
\times 1$ vector ${\bf x}_T[i] = \big[ {\bf x}_1^T[i] ~ {\bf
x}_2^T[i] ~ \ldots ~ {\bf x}_{N_R}^T[i] \big]^T$ is composed by
the data symbols transmitted from the $N_T$ antennas at the
transmitter with ${\bf x}_{k}[i]$ being the $i$th transmitted
block with dimensions $L \times 1$. The $L N_R \times 1$ vector
${\bf n}[i]$ is a complex Gaussian noise vector with zero mean and
$E \big[ {\bf n}[i] {\bf n}^H[i] \big] = \sigma^2 {\bf I}$, where
$(\cdot)^{T}$ and $(\cdot)^{H}$ denote transpose and Hermitian
transpose, respectively, and $E[\cdot]$ stands for expected value.

\section{Adaptive MIMO Decision Feedback Equalizer}

\begin{figure}[h!]
  \begin{center}
 \def\eepicsize#1#2{1\columnwidth}
 \input{./fig2.eepic}
    \caption{MIMO system model with conventional decision feedback equalizer.}
    \label{fig:dfesystem}
  \end{center}
\end{figure}
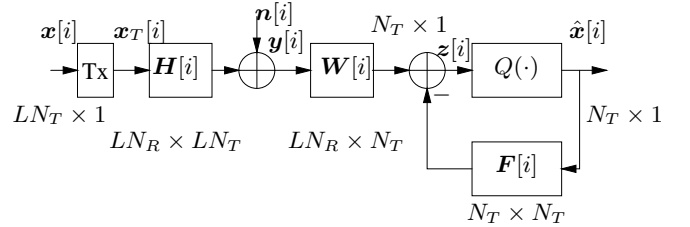

The conventional MIMO decision feedback equalizer design
corresponds to determining an $L N_R \times N_T$ estimator ${\bf
W}[i] = \big[ {\bf w}_1[i] {\bf w}_2[i] \ldots {\bf w}_{N_T}[i]
\big]$ that linearly combines the received signal ${\bf y}[i]$ and
an $N_T \times N_T$ estimator ${\bf F}[i] = \big[ {\bf f}_1[i]~
{\bf f}_2[i]~ \ldots ~{\bf f}_{N_T}[i] \big]$ that cancel the
associated MAI created by the different data streams and the ISI
caused by the multipath channels. The block diagram shown in Fig.
2 illustrates how the MIMO DFE works. The estimate ${\bf z}[i]$ of
the desired symbols is given by
\begin{equation}
{\bf z} [i] = {\bf W}^H[i] {\bf y}[i] - {\bf F}^{H}[i] \hat{\bf
x}[i],
\end{equation}
where $\hat{\bf x}[i] = Q \big( {\bf W}^H[i] {\bf y}[i] \big)$ is
the initial decision vector taken with the feedforward section
${\bf W}[i]$ and $Q \big( \cdot \big)$ represents a decision
device. For the design of the feedback section, we constrain ${\bf
F}[i]$ to be full and to have zeros along the diagonal to avoid
cancelling the desired symbols. This corresponds to parallel
decision feedback \cite{sun,woodward2}. Specifically, the non-zero
part of the filter ${\bf F}[i]$ corresponds to the number of used
feedback connections and to the users to be cancelled. The
feedback connections used and their associated number of non-zero
filter coefficients in ${\bf F}[i]$ are equal to $N_T-1$ for all
users.

The detected symbols of the DFE after the interference
cancellation $\hat{\bf x}^{(f)}[i]$ carried out by the feedback
section are given by
\begin{equation}
\hat{\bf x}^{(f)}[i] = Q \big( {\bf z}[i] \big).
\end{equation}
In order to design the $L N_R$-dimensional estimators ${\bf
w}_j[i]$ ($j=1, \ldots, N_T$) that form ${\bf W}[i]$, one can
resort to stochastic gradient or LS algorithms, as in
\cite{tehrani,dhahir,kominakis}.

\section{Proposed Adaptive MIMO Reduced-Rank Decision Feedback Equalizer}

\begin{figure}[h!]
  \begin{center}
    \def\eepicsize#1#2{1\columnwidth} \input{./fig3.eepic}
    \caption{Proposed MIMO reduced-rank decision feedback equalizer.}
    \label{fig:propdfesystem}
  \end{center}
\end{figure}
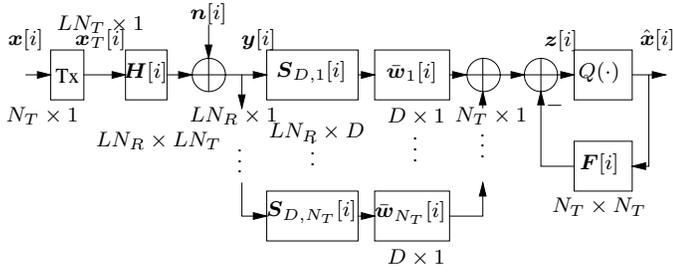

In the proposed reduced-rank linear MIMO equalizer, the signal
processing tasks are carried out in two stages, as illustrated in
Fig. 3. Firstly, we consider the dimensionality reduction of ${\bf
y}[i]$ by projecting the received vector onto a lower dimensional
subspace. Specifically, consider an $L N_R \times D$ projection
matrix ${\bf S}_{D,j}[i]$ which carries out a dimensionality
reduction on the received data for extracting the symbols
transmitted from antenna $j$ as given by
\begin{equation}
\bar{\bf y}_j[i] = {\bf S}_{D,j}^H[i] {\bf y}[i],
\end{equation}
where, in what follows, all $D$-dimensional quantities are denoted
with a "bar". The resulting projected received vector $\bar{\bf
y}_j[i]$ is the input to an estimator represented by the $D \times
1$ vector $\bar{\bf w}_j[i]=[ \bar{w}_{j,1}^{}[i]
~\bar{w}_{j,2}^{}[i]~\ldots\bar{w}_{j,D}^{}[i]]^T$ for time
interval $i$. The reduced-rank estimator output corresponding to
the $i$th time instant and estimators $\bar{\bf w}_j[i]$, ${\bf
S}_{D,j}[i]$, and ${\bf f}_j[i]$ for extracting the symbol
transmitted from antenna $j$ is
\begin{equation}
\begin{split}
z_j[i] & = \bar{\bf w}^{H}_j[i] {\bf S}_{D,j}^H[i] {\bf y}[i] -
{\bf f}_j^H[i] \hat{\bf x}[i] \\ & = \bar{\bf w}^{H}_j[i]\bar{\bf
y}[i] - {\bf f}_j^H[i] \hat{\bf x}[i],
\end{split}
\end{equation}
where the feedback section estimators form ${\bf F}[i] = \big[
{\bf f}_1[i] ~ \ldots ~ {\bf f}_{N_T}[i] \big]$, which is
constrained to have zeros along the main diagonal to avoid
cancelling the desired symbols, similarly to the conventional MIMO
DFE. We will consider here the case where the number of transmit
antennas $N_T$ is reasonably small which leads to small estimators
in ${\bf F}[i]$.
From the above outputs, we construct the vector ${\bf z}[i] =
\big[ z_1[i] \ldots z_j[i] \ldots z_{N_T}[i] \big]^T$. The
detected symbols of the proposed reduced-rank MIMO DFE after the
interference cancellation $\hat{\bf x}^{(f)}[i]$ are obtained with
\begin{equation}
\hat{\bf x}^{(f)}[i] = Q \big( {\bf z}[i] \big).
\end{equation}

We remark that when $N_T$ becomes large and consequently the
number of used feedback connections corresponding to the data
streams to be cancelled, the designer can incorporate this in a
single large matrix filter ${\bf C}^T[i] = \big[ {\bf W}^T[i] {\bf
F}^T[i] \big] = \big[{\bf c}_1^T[i] ~ \ldots ~ {\bf c}_{N_T}^T
\big]$ and stack the vectors ${\bf y}[i]$ and $\hat{\bf x}[i]$,
forming the $(LN_R + N+T) \times 1$ vector ${\bf a}^T [i] = \big[
{\bf y}^T[i] \ms \hat{\bf x}^T[i] \big]$. In this case, the output
of the MIMO DFE would be ${z}_j[i] = {\bf c}^H_j[i] \tilde{\bf
S}_{D,j}^H[i]{\bf a}[i]$, where $\tilde{\bf S}_{D,j}[i]$ is a
modified projection matrix with dimensions $(LN_R + N+T) \times
D$.

\section{Proposed Least Squares Design and Reduced-Rank RLS Algorithms}

In this section, we present a joint iterative exponentially
weighted least squares (LS) estimator design of the parameters
${\bf S}_{D,j}[i]$, ${\bf w}_j[i]$, and ${\bf f}_j[i]$ of the
proposed MIMO reduced-rank DFE and a computationally efficient RLS
algorithm for implementing the proposed LS estimator.

\subsection{Least Squares Design}

In order to design ${\bf S}_{D,j}[i]$, $\bar{\bf w}_j[i]$, and
${\bf f}_j[i]$, we describe a joint iterative LS optimization
algorithm. Let us consider the exponentially-weighted LS
expressions for the estimators ${\bf S}_{D,j}[i]$ , $\bar{\bf
w}_j[i]$, and ${\bf f}_j[i]$ can be computed via the cost function
\begin{equation}
\begin{split}
{\mathcal{C}} & = \sum_{l=1}^i \lambda^{i-l} \big| x_j[l] -
\bar{\bf w}^H_j[i] {\bf S}_{D,j}^H[i] {\bf y}[l] +{\bf f}^H_{j}[i]
\hat{\bf x}[l] \big|^2,
\end{split}
\end{equation}
where $\lambda$ is the forgetting factor.

By minimizing (8) with respect to ${\bf S}_{D,j}[i]$, we obtain
\begin{equation} {\bf S}_{D,j}[i] = {\bf R}^{-1}[i] \Big( {\bf P}_{D,j}[i] + {\bf P}_{{\bf f}_{j}}[i] \Big) {\bf
R}_{\bar{{\bf w}}_j}^{-1}[i],
\end{equation}
where ${\bf P}_{D,j}[i] = \sum_{l=1}^i \lambda^{i-l}x^{*}_j[l]{\bf
y}[l]{\bf w}^H_j[i]$, ${\bf R}[i] = \sum_{l=1}^i \lambda^{i-l}{\bf
y}[l]{\bf y}^{H}[l]$, ${\bf P}_{{\bf f}_{j}}[i] = \sum_{l=1}^i
\lambda^{i-l} {\bf y}[l] {\bf w}_j^H [i] \hat{\bf x}^H[l]$, and
${\bf R}_{w_j}[i] = \sum_{l=1}^i \lambda^{i-l}{\bf w}_j[l]{\bf
w}^{H}_j[l]$. By minimizing (8) with respect to $\bar{\bf
w}_j[i]$, the reduced-rank estimator becomes
\begin{equation}
\bar{\bf w}_j[i] = \bar{\bf R}^{-1}_j[i] \Big( \bar{\bf p}_j[i] +
{\bf D}_j[i] {\bf f}_j[i] \Big),
\end{equation}
where $\bar{\bf p}_j[i] = {\bf S}_{D,j}^H[i] \sum_{l=1}^i
\lambda^{i-l}x^{*}_j[l]{\bf y}[l] = \sum_{l=1}^i
\lambda^{i-l}x^{*}_j[l]\bar{\bf y}[l]]$, the reduced-rank
estimated correlation matrix is $\bar{\bf R}_j[i] = {\bf
S}_{D,j}^H[i]\sum_{l=1}^i \lambda^{i-l}{\bf y}[l]{\bf y}^H[l] {\bf
S}_{D,j}[i] $ and ${\bf D}_j[i] = {\bf S}_{D,j}^H[i] \sum_{l=1}^i
\lambda^{i-l} {\bf y}[l] \hat{\bf x}^H[l]$.  By minimizing (8)
with respect to ${\bf f}_{j}[i]$ we obtain
\begin{equation} {\bf f}_{j}[i] = {\bf B}^{-1}[i] \Big( {\bf D}_{j}^H[i]
\bar{\bf w}_j[i] - {\bf b}[i] \Big),
\end{equation}
where ${\bf B}[i] = \sum_{l=1}^i \lambda^{i-l}\hat{\bf
x}[l]\hat{\bf x}^{H}[l]$ and ${\bf b}[i] = \sum_{l=1}^i
\lambda^{i-l} x_j^*[l]\hat{\bf x}[l]$. The associated sum of error
squares (SES) is
\begin{equation}
\begin{split}
{\rm SES} & = \sigma^{2}_{x}  - \bar{\bf w}_j^H[i] {\bf
S}_{D,j}^H[i] {\bf p}[i] - {\bf p}^H[i]{\bf S}_{D,j}[i]\bar{\bf
w}_j[i] \\ & \quad + \bar{\bf w}^H[i] {\bf S}_{D,j}^H[i] {\bf
R}{\bf S}_{D,j}[i] \bar{\bf w}_j[i] \\ & \quad -
{\bf f}_j^H[i] {\bf D}_j^H[i] {\bf S}_{D,j}[i] \bar{\bf w}_j[i] \\
& \quad - \bar{\bf w}^H[i] {\bf S}_{D,j}[i] {\bf D}_j[i] {\bf
f}_j[i]  + {\bf f}_j^H{\bf b}[i] \\ & \quad + {\bf b}^[i]{\bf
f}_j[i] + {\bf f}_j^H[i] {\bf B}[i] {\bf f}_j[i],
\end{split}
\end{equation}
where $\sigma^{2}_{x}=\sum_{l=1}^i \lambda^{i-l}|x[l]|^{2}$. Note
that the expressions in (9), (10), and (11) are not closed-form
solutions for $\bar{\bf w}_j[i]$, ${\bf S}_{D,j}[i]$, and ${\bf
f}_j[i]$ since they depend on each other and, thus, they have to
be iterated with an initial guess to obtain a solution. The key
strategy lies in the joint optimization of the estimators. The
rank $D$ must be set by the designer to ensure appropriate
performance. The computational complexity of implementing (9),
(10), and (11) is cubic with the number of elements in the
estimators, namely, $LN_R$, $D$, and $N_T$, respectively. In what
follows, we introduce efficient RLS algorithms for implementing
the estimators with a quadratic cost.

\subsection{Reduced-Rank Recursive Least Squares Algorithms}

In this part, we present RLS algorithms for efficiently
implementing the LS design of the previous subsection. Firstly,
let us define ${\bf
  P}[i] ={\bf R}^{-1}[i]$, ${\bf Q}_{\bar{\bf w}_j}[i-1]={\bf
  P}^{-1}_{\bar{\bf w}_j}[i]$, ${\bf P}_{D,j} [i] = \lambda {\bf
  P}_{D}[i-1] + x^*_j[i] {\bf y}[i] {\bf w}^H [i]$, ${\bf P}_{f,j} [i] =
\lambda {\bf P}_{f,j}[i-1] + x^*_j[i] {\bf y}[i] \bar{\bf w}^H
[i]$, and rewrite the expression in (9) as follows
\begin{equation}
\begin{split}
{\bf S}_{D,j} [i] & = {\bf R}^{-1}[i] \big( {\bf P}_{D,j} [i] +
{\bf P}_{{\bf f}_j}[i] \big) {\bf P}_{\bar{\bf w}_j}[i] \\
& = {\bf P}[i] \big( {\bf P}_{D,j} [i] +
{\bf P}_{{\bf f}_j}[i] \big) {\bf Q}_{\bar{\bf w}_j}[i-1] \\
& = {\bf S}_{D,j}[i-1] + {\bf k}[i] \big( { x}_j^*[i] {\bf
t}_j^H[i] \\ & \quad - {\bf y}^H[i]{\bf S}_{D,j}[i-1] + {\bf
t}_j^H[i] \hat{\bf x}^H[i] {\bf f}_j[i]  \big).
\end{split}
\end{equation}
By defining the vector ${\bf t}_j[i] = {\bf Q}_{\bar{\bf w}_j}[i]
\bar{\bf w}_j[i]$ and using the fact that $\bar{\bf y}^H[i-1] =
{\bf y}^H[i-1]{\bf S}_{D,j}[i-1]$ we arrive at
\begin{equation}
\begin{split}
{\bf S}_{D,j}[i] & = {\bf S}_{D,j}[i-1] - {\bf k}[i] \big(
x^*_j[i] {\bf t}^H[i] \\ & \quad - {\bf y}^H[i]{\bf S}_{D,j}[i-1]
+ {\bf t}_j^H[i] \hat{\bf x}^H[i] {\bf f}_j[i] \big),
\end{split}
\end{equation}
where the Kalman gain vector for the estimation of ${\bf
S}_{D,j}[i]$ is
\begin{equation}
{\bf k}[i] = \frac{\lambda^{-1} {\bf P}[i-1] {\bf y}[i] }{1 +
\lambda^{-1} {\bf y}^H[i] {\bf P}[i-1]{\bf y}[i]},
\end{equation}
and the update for the matrix ${\bf P}[i]$ employs the matrix
inversion lemma
\begin{equation}
{\bf P}[i] = \lambda^{-1} {\bf P}[i-1] - \lambda^{-1} {\bf k}[i]
{\bf y}^H[i] {\bf P}[i-1],
\end{equation}
the vector ${\bf t}_j[i]$ is updated as follows
\begin{equation}
{\bf t}_j[i] = \frac{\lambda^{-1} {\bf Q}_{\bar{\bf w}_j}[i-1]
\bar{\bf w}_j[i-1] }{1 + \lambda^{-1} \bar{\bf w}^H_j[i-1] {\bf
Q}_{\bar{\bf w}_j}[i-1]\bar{\bf w}_j[i-1]},
\end{equation}
and the matrix inversion lemma is used to update ${\bf
Q}_{\bar{\bf w}_j}[i]$ as described by
\begin{equation}
{\bf Q}_{\bar{\bf w}_j}[i] = \lambda^{-1} {\bf Q}_{\bar{\bf
w}_j}[i-1] - \lambda^{-1} {\bf t}_j[i] \bar{\bf w}^H_j[i-1].
\end{equation}
Equations (13)-(18) constitute the part of the proposed RLS
algorithms for estimating the projection matrix ${\bf
S}_{D,j}[i]$. In order to develop the second part of the algorithm
that estimates $\bar{\bf
  w}_j[i]$, let us consider the expression in (10) with its associated
quantities, i.e., the matrix $\bar{\bf R}_j[i] = \sum_{l=1}^i
\lambda^{i-l}\bar{\bf y}[l] \bar{\bf y}^{H}[l]$, the vector
$\bar{\bf
  p}_j[i] = \sum_{l=1}^i \lambda^{i-l}x^{*}_j[l]\bar{\bf y}[l]$ and
the matrix ${\bf D}_j[i] = \sum_{l=1}^i \lambda^{i-l}\bar{\bf
  y}[l]\hat{\bf x}^{H}[l]$. Let us define
$\boldsymbol{\bar{\Phi}_j}[i] = {\bf R}^{-1}_j[i]$ and rewrite
$\bar{\bf p}_j[i]$ as $\bar{\bf p}_j[i] = \lambda \bar{\bf
p}_j[i-1] + x^*_j[i] \bar{\bf y}_j[i]$ and ${\bf D}_j[i]$ as ${\bf
D}_j[i] = {\bf
  D}_j[i-1] + \bar{\bf y}[i]\hat{\bf x}^{H}[i]$. We can write (10) as
follows
\begin{equation}
\begin{split}
\bar{\bf w}_j[i] & = \boldsymbol{\bar{\Phi}_j}[i] \big( \bar{\bf
p}_{j}[i] + {\bf D}_j[i] {\bf f}_j[i] \big) \\
& = \bar{\bf w}_j[i-1] - \bar{\bf k}_j[i] \bar{\bf y}^H_j[i]
\bar{\bf w}_j[i-1]  + \bar{\bf k}_j[i] x^*_j[i] \\ & \quad + \hat{\bf x}^H[i] {\bf f}_j[i]  \\
& = \bar{\bf w}[i-1] + \bar{\bf k}_j[i] \big( x^*_j[i] - \bar{\bf
y}^H_j[i] \bar{\bf w}_j[i-1] + \hat{\bf x}^H[i] {\bf f}_j[i]\big).
\end{split}
\end{equation}
By defining $\xi_j[i] = x_j[i] - \bar{\bf w}^H_j[i-1]\bar{\bf
y}_j[i] + {\bf f}_j^H[i] \hat{\bf x}[i]$ we arrive at the proposed
RLS algorithm for estimating $\bar{\bf w}_j[i]$
\begin{equation}
\bar{\bf w}_j[i] = \bar{\bf w}_j[i-1] + \bar{\bf k}_j[i]
\xi^*_j[i],
\end{equation}
where the Kalman gain vector is given by
\begin{equation}
\bar{\bf k}_j[i] = \frac{\lambda^{-1}
\boldsymbol{\bar{\Phi}_j}[i-1] \bar{\bf y}_j[i] }{1 + \lambda^{-1}
\bar{\bf y}^H_j[i] \boldsymbol{\bar{\Phi}_j}[i-1]\bar{\bf
y}_j[i]},
\end{equation}
and the update for the matrix inverse $\boldsymbol{\bar{\Phi}}[i]$
employs the matrix inversion lemma
\begin{equation}
\boldsymbol{\bar{\Phi}_j}[i] = \lambda^{-1}
\boldsymbol{\bar{\Phi}_j}[i-1] - \lambda^{-1} \bar{\bf k}_j[i]
\bar{\bf y}^H_j[i] \boldsymbol{\bar{\Phi}_j}[i-1].
\end{equation}
The third part of the proposed algorithm estimates the feedback
filter section ${\bf f}_j[i]$ and uses the expression in (11) in a
similar way to the development of the estimation procedure of
$\bar{{\bf w}}_j[i]$. Let us define ${\bf P}_B[i] = {\bf
B}^{-1}[i]$ and rewrite ${\bf b}[i]$ as ${\bf b}[i] = \lambda{\bf
b}[i-1] + x^*_j[i] \bar{\bf x}[i]$. Then, let us express (11) in
an alternative way as follows
\begin{equation}
\begin{split}
{\bf f}_j[i] & = {\bf P}_B [i] \big( {\bf D}_j^H[i] \bar{\bf
w}_j[i] - {\bf b}[i] \big)  \\
& = {\bf f}_j[i-1] - {\bf k}_{B,j}[i] \big( x_j^*[i] - \bar{\bf
y}_j^H[i] \bar{\bf w}_j[i] + \hat{\bf x}^H[i]{\bf f}_j[i-1] \big)
\\ & = {\bf f}[i-1] - {\bf k}_{B,j}[i] \xi_k^*[i],
\end{split}
\end{equation}
where the Kalman gain vector ${\bf k}_B[i]$ is given by
\begin{equation}
{\bf k}_B[i] = \frac{\lambda^{-1} {\bf P}_B[i-1] \hat{\bf x}[i]
}{1 + \lambda^{-1} \hat{\bf x}^H[i] {\bf P}_B[i-1] \hat{\bf
x}[i]},
\end{equation}
and the update for the matrix inverse ${\bf P}_B[i]$ employs the
matrix inversion lemma
\begin{equation}
{\bf P}_B[i] = \lambda^{-1} {\bf P}_B[i-1] - \lambda^{-1} {\bf
k}_B[i] \hat{\bf x}^H[i] {\bf P}_B[i-1].
\end{equation}
Equations (19)-(25) constitute the second and the third part of
the proposed algorithm, which estimate $\bar{\bf w}_j[i]$ and
${\bf
  f}_j[i]$. The computational complexity of the proposed RLS
algorithms is $O(D^2)$ for the estimation of $\bar{\bf w}_j[i]$,
$O((LN_R)^2)$ for the estimation of ${\bf S}_{D,j}[i]$ and
$O(N_T^2)$ for the estimation of ${\bf f}_j[i]$. Because $D <<
LN_R$, as will be explained in the next section, the overall
complexity is in the same order of the conventional full-rank RLS
algorithm ($O((LN_R)^2)$) .

\section{Simulations}

In this section, we assess and compare the bit error rate~(BER)
performance of the adaptive MIMO decision feedback equalization
schemes with different estimators designed according to the LS
criterion, namely, the full-rank \cite{tehrani}, the reduced-rank
MWF \cite{goldstein} when the turbo coding and decoding of
\cite{sun} is removed, and the AVF \cite{avf} techniques for the
design of the receivers. The proposed adaptive reduced-rank MIMO
DFE employs a reduced-rank estimator $\bar{\bf w}_j[i]$ with $D$
coefficients for the feedforward section, followed by a feedback
structure estimator ${\bf f}_j[i]$ with full-rank estimators that
perform interference cancellation. For all simulations, we use the
initial values $\bar{\bf w}_j[0]= \big[ 1, 0, \ldots, 0 \big]$,
${\bf f}_{j} = {\bf 0}_{N_T\times 1}$, and ${\bf S}_{D,j}[0]=[{\bf
I}_D ~ {\bf 0}_{D \times (JM-D)}]^T$, assume $L=5$ as an upper
bound, use $3$-path channels spaced by one symbol and with
relative powers taken from complex Gaussian random variables with
zero mean and unit variance and QPSK modulation. The channel is
static over one packet, we average the experiments over $100$ runs
and define the signal-to-noise ratio (SNR) as $\textrm{SNR} = 10
\log_{10} \frac{N_T \sigma_x^2}{\sigma^2}$, where $\sigma_x^2$ is
the variance of the transmitted symbols and $\sigma^2$ is the
noise variance. The adaptive MIMO DFE employs $N_T=4$ and $N_R=8$
in a spatial multiplexing configuration, leading to estimators at
the receiver with $L N_R = 40$ elements. The adaptive LS
estimators of all methods are trained with $250$ symbols and then
switched to decision-directed mode.

We consider the BER performance versus the rank $D$ with optimized
parameters (forgetting factors $\lambda=0.998$) for all schemes.
The results in Fig. 4 indicate that the best rank for the proposed
scheme is $D=4$ (which will be used in the remaining experiments)
and it is close to the optimal MMSE that assumes the knowledge of
the channel and the noise variance. Our studies with systems with
different sizes show that the optimal rank $D$ does not vary
significantly with the system size. It remains in a small range of
values and does not scale with system size, which brings
considerably faster convergence speed. However, It should also be
remarked that the optimal rank $D$ depends on the data record size
and other parameters of the systems. In order to tackle this
problem, we plan to devise in the future an adaptive rank
selection algorithm that will automatically adjust the best rank
and will be considered elsewhere.

\begin{figure}[!htb]
\begin{center}
\def\epsfsize#1#2{1\columnwidth}
\epsfbox{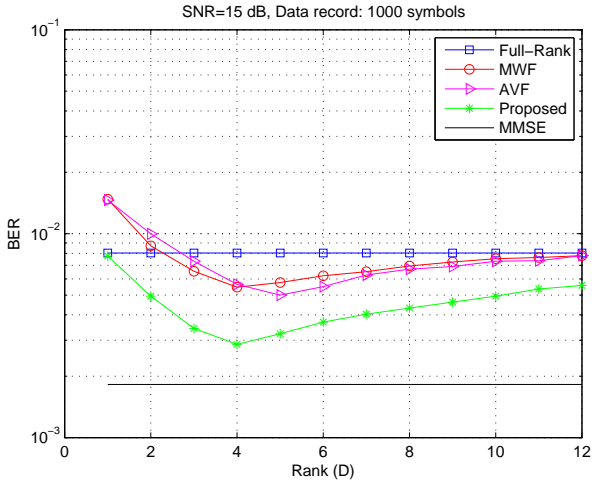} \caption{BER performance versus rank (D).}
\end{center}
\end{figure}

The BER convergence performance versus number of received symbols
is shown in Fig. 5. The results show that the proposed scheme has
a significantly faster convergence performance and obtains good
gains over the best known schemes. The BER performance versus the
signal-to-noise ratio (SNR) is shown in Fig. 6. The plots show
that the proposed reduced-rank MIMO equalizer and estimation
algorithm have the best performance followed by the AVF, the MWF,
and the full-rank estimator.

\begin{figure}[!htb]
\begin{center}
\def\epsfsize#1#2{1\columnwidth}
\epsfbox{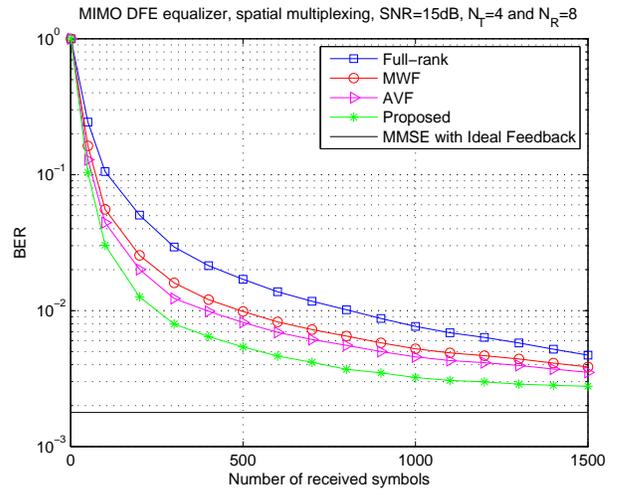} \caption{BER performance versus number of
received symbols.}
\end{center}
\end{figure}

The advantages of the reduced-rank estimators are due to the
reduced amount of training and the relatively short data record
(packet size). Therefore, for packets with relatively small size,
the faster training of reduced-rank LS estimators will lead to
superior BER to conventional full-rank LS estimators. As the
length of the packets is increased, the advantages of reduced-rank
estimators become less pronounced for training purposes and so
become the BER advantages over full-rank estimators. In comparison
with the MWF and AVF reduced-rank schemes, the proposed scheme
exploits the joint and iterative exchange of information between
the projection matrix and the reduced-rank estimators, which leads
to better performance.

\begin{figure}[!htb]
\begin{center}
\def\epsfsize#1#2{1\columnwidth}
\epsfbox{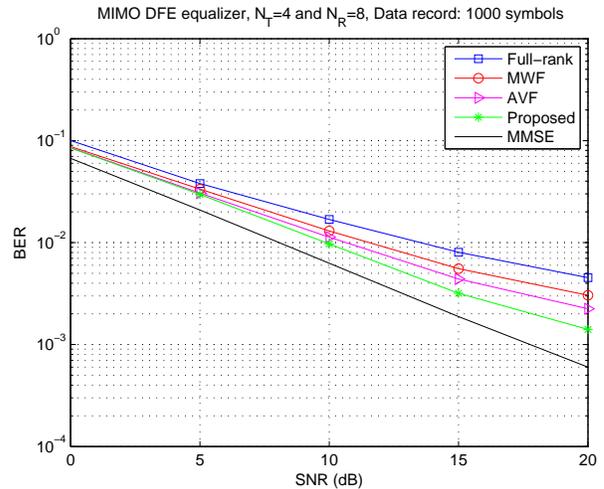} \caption{BER performance versus $E_b/N_0$ .}
\end{center}
\end{figure}

\section{Concluding Remarks}

This paper proposed an adaptive reduced-rank MIMO decision
feedback equalization (DFE) structure based on joint iterative
optimization of adaptive estimators. We described LS expressions
and efficient RLS algorithms for the design of the proposed MIMO
DFE. Simulations for a MIMO equalization application show that the
proposed schemes outperforms the state-of-the-art reduced-rank and
the conventional estimation algorithms at about the same
complexity. Future work will consider the application of the
proposed  MIMO DFE equalizers and reduced-rank
estimators to realistic MIMO channel models. \\

\end{document}

%% file: fig1.eepic
\setlength{\unitlength}{0.00062500in}
\begingroup\makeatletter\ifx\SetFigFont\undefined%
\gdef\SetFigFont#1#2#3#4#5{%
  \reset@font\fontsize{#1}{#2pt}%
  \fontfamily{#3}\fontseries{#4}\fontshape{#5}%
  \selectfont}%
\fi\endgroup%
{\renewcommand{\dashlinestretch}{30}
\begin{picture}(5186,2673)(0,-10)
\put(224,1558){\makebox(0,0)[b]{{\SetFigFont{9}{10.8}{\rmdefault}{\mddefault}{\updefault}$x[i]$}}}
\path(974,433)(1199,433)(1199,733)
	(1349,883)(1049,883)(1199,733)
\path(3224,1521)(2961,1521)(2961,1821)
	(3111,1971)(2811,1971)(2961,1821)
\path(3224,433)(2999,433)(2999,733)
	(3149,883)(2849,883)(2999,733)
\path(3224,2196)(2961,2196)(2961,2496)
	(3111,2646)(2811,2646)(2961,2496)
\path(974,2158)(1199,2158)(1199,2458)
	(1349,2608)(1049,2608)(1199,2458)
\path(1424,2608)(1949,2233)(2099,2533)(2624,2308)
\blacken\path(2501.885,2327.696)(2624.000,2308.000)(2525.520,2382.845)(2501.885,2327.696)
\path(1424,1933)(1949,1558)(2099,1858)(2624,1633)
\blacken\path(2501.885,1652.696)(2624.000,1633.000)(2525.520,1707.845)(2501.885,1652.696)
\path(1424,883)(1949,508)(2099,808)(2624,583)
\blacken\path(2501.885,602.696)(2624.000,583.000)(2525.520,657.845)(2501.885,602.696)
\path(449,2383)(974,2383)(974,358)
	(449,358)(449,2383)
\path(3224,2383)(3749,2383)(3749,358)
	(3224,358)(3224,2383)
\path(3749,1408)(4349,1408)
\blacken\path(4229.000,1378.000)(4349.000,1408.000)(4229.000,1438.000)(4229.000,1378.000)
\path(4349,2383)(4874,2383)(4874,358)
	(4349,358)(4349,2383)
\path(4874,1408)(5174,1408)
\blacken\path(5054.000,1378.000)(5174.000,1408.000)(5054.000,1438.000)(5054.000,1378.000)
\path(74,1408)(449,1408)
\blacken\path(329.000,1378.000)(449.000,1408.000)(329.000,1438.000)(329.000,1378.000)
\put(1274,58){\makebox(0,0)[b]{{\SetFigFont{9}{10.8}{\rmdefault}{\mddefault}{\updefault}$N_{T}$}}}
\put(2924,58){\makebox(0,0)[b]{{\SetFigFont{9}{10.8}{\rmdefault}{\mddefault}{\updefault}$N_{R}$}}}
\put(1199,1183){\makebox(0,0)[b]{{\SetFigFont{9}{10.8}{\rmdefault}{\mddefault}{\updefault}$\vdots$}}}
\put(2999,1183){\makebox(0,0)[b]{{\SetFigFont{9}{10.8}{\rmdefault}{\mddefault}{\updefault}$\vdots$}}}
\put(3449,1408){\makebox(0,0)[b]{{\SetFigFont{9}{10.8}{\rmdefault}{\mddefault}{\updefault}Rx}}}
\put(4049,1558){\makebox(0,0)[b]{{\SetFigFont{9}{10.8}{\rmdefault}{\mddefault}{\updefault}$z[i]$}}}
\put(4574,1408){\makebox(0,0)[b]{{\SetFigFont{9}{10.8}{\rmdefault}{\mddefault}{\updefault}$Q(\cdot)$}}}
\put(5099,1558){\makebox(0,0)[b]{{\SetFigFont{9}{10.8}{\rmdefault}{\mddefault}{\updefault}$\hat{x}[i]$}}}
\put(674,1408){\makebox(0,0)[b]{{\SetFigFont{9}{10.8}{\rmdefault}{\mddefault}{\updefault}Tx}}}
\path(974,1558)(1199,1558)(1199,1858)
	(1349,2008)(1049,2008)(1199,1858)
\end{picture}
}

%% file: fig2.eepic
\setlength{\unitlength}{0.00062500in}
\begingroup\makeatletter\ifx\SetFigFont\undefined%
\gdef\SetFigFont#1#2#3#4#5{%
  \reset@font\fontsize{#1}{#2pt}%
  \fontfamily{#3}\fontseries{#4}\fontshape{#5}%
  \selectfont}%
\fi\endgroup%
{\renewcommand{\dashlinestretch}{30}
\begin{picture}(4812,1879)(0,-10)
\put(1986,1474){\makebox(0,0)[b]{{\SetFigFont{9}{10.8}{\rmdefault}{\mddefault}{\updefault}$\boldsymbol{y}[i]$}}}
\put(3162,1333){\ellipse{300}{300}}
\path(12,1333)(237,1333)
\blacken\path(117.000,1303.000)(237.000,1333.000)(117.000,1363.000)(117.000,1303.000)
\path(3312,1333)(3537,1333)
\blacken\path(3417.000,1303.000)(3537.000,1333.000)(3417.000,1363.000)(3417.000,1303.000)
\path(4287,1333)(4662,1333)
\blacken\path(4542.000,1303.000)(4662.000,1333.000)(4542.000,1363.000)(4542.000,1303.000)
\path(4437,1333)(4437,508)(4287,508)
\blacken\path(4407.000,538.000)(4287.000,508.000)(4407.000,478.000)(4407.000,538.000)
\path(237,1558)(537,1558)(537,1108)
	(237,1108)(237,1558)
\path(537,1333)(837,1333)
\blacken\path(717.000,1303.000)(837.000,1333.000)(717.000,1363.000)(717.000,1303.000)
\path(837,1558)(1362,1558)(1362,1108)
	(837,1108)(837,1558)
\path(1362,1333)(1587,1333)
\blacken\path(1467.000,1303.000)(1587.000,1333.000)(1467.000,1363.000)(1467.000,1303.000)
\path(1587,1333)(1887,1333)
\path(1887,1333)(2187,1333)
\blacken\path(2067.000,1303.000)(2187.000,1333.000)(2067.000,1363.000)(2067.000,1303.000)
\path(1737,1483)(1737,1183)
\path(1737,1783)(1737,1483)
\blacken\path(1707.000,1603.000)(1737.000,1483.000)(1767.000,1603.000)(1707.000,1603.000)
\path(2187,1558)(2712,1558)(2712,1108)
	(2187,1108)(2187,1558)
\path(2703,1333)(3012,1333)
\blacken\path(2892.000,1303.000)(3012.000,1333.000)(2892.000,1363.000)(2892.000,1303.000)
\path(3162,1483)(3162,1183)
\path(3537,508)(3162,508)(3162,1183)
\blacken\path(3192.000,1063.000)(3162.000,1183.000)(3132.000,1063.000)(3192.000,1063.000)
\path(3537,733)(4287,733)(4287,283)
	(3537,283)(3537,733)
\path(3537,1558)(4287,1558)(4287,1108)
	(3537,1108)(3537,1558)
\path(3012,1333)(3087,1333)(3162,1333)
	(3237,1333)(3312,1333)
\put(387,1258){\makebox(0,0)[b]{{\SetFigFont{9}{10.8}{\rmdefault}{\mddefault}{\updefault}Tx}}}
\put(87,1558){\makebox(0,0)[b]{{\SetFigFont{9}{10.8}{\rmdefault}{\mddefault}{\updefault}$\boldsymbol{x}[i]$}}}
\put(762,1558){\makebox(0,0)[b]{{\SetFigFont{9}{10.8}{\rmdefault}{\mddefault}{\updefault}$\boldsymbol{x}_{T}[i]$}}}
\put(1062,1258){\makebox(0,0)[b]{{\SetFigFont{9}{10.8}{\rmdefault}{\mddefault}{\updefault}$\boldsymbol{H}[i]$}}}
\put(2487,1258){\makebox(0,0)[b]{{\SetFigFont{9}{10.8}{\rmdefault}{\mddefault}{\updefault}$\boldsymbol{W}[i]$}}}
\put(3912,1258){\makebox(0,0)[b]{{\SetFigFont{9}{10.8}{\rmdefault}{\mddefault}{\updefault}$Q(\cdot)$}}}
\put(4512,1558){\makebox(0,0)[b]{{\SetFigFont{9}{10.8}{\rmdefault}{\mddefault}{\updefault}$\hat{\boldsymbol{x}}[i]$}}}
\put(4812,883){\makebox(0,0)[b]{{\SetFigFont{9}{10.8}{\rmdefault}{\mddefault}{\updefault}$N_{T}\times 1$}}}
\put(3012,1633){\makebox(0,0)[b]{{\SetFigFont{9}{10.8}{\rmdefault}{\mddefault}{\updefault}$N_{T}\times 1$}}}
\put(2487,658){\makebox(0,0)[b]{{\SetFigFont{9}{10.8}{\rmdefault}{\mddefault}{\updefault}$LN_{R}\times N_{T}$}}}
\put(1062,658){\makebox(0,0)[b]{{\SetFigFont{9}{10.8}{\rmdefault}{\mddefault}{\updefault}$LN_{R}\times LN_{T}$}}}
\put(87,883){\makebox(0,0)[b]{{\SetFigFont{9}{10.8}{\rmdefault}{\mddefault}{\updefault}$LN_{T}\times 1$}}}
\put(3912,433){\makebox(0,0)[b]{{\SetFigFont{9}{10.8}{\rmdefault}{\mddefault}{\updefault}$\boldsymbol{F}[i]$}}}
\put(3912,58){\makebox(0,0)[b]{{\SetFigFont{9}{10.8}{\rmdefault}{\mddefault}{\updefault}$N_{T}\times N_{T}$}}}
\put(3387,1408){\makebox(0,0)[b]{{\SetFigFont{9}{10.8}{\rmdefault}{\mddefault}{\updefault}$\boldsymbol{z}[i]$}}}
\put(1887,1708){\makebox(0,0)[b]{{\SetFigFont{9}{10.8}{\rmdefault}{\mddefault}{\updefault}$\boldsymbol{n}[i]$}}}
\put(3285,1060){\makebox(0,0)[b]{{\SetFigFont{9}{10.8}{\rmdefault}{\mddefault}{\updefault}$-$}}}
\put(1737,1333){\ellipse{300}{300}}
\end{picture}
}

%% file: fig3.eepic
\setlength{\unitlength}{0.00057500in}
\begingroup\makeatletter\ifx\SetFigFont\undefined%
\gdef\SetFigFont#1#2#3#4#5{%
  \reset@font\fontsize{#1}{#2pt}%
  \fontfamily{#3}\fontseries{#4}\fontshape{#5}%
  \selectfont}%
\fi\endgroup%
{\renewcommand{\dashlinestretch}{30}
\begin{picture}(5799,2395)(0,-10)
\put(2588,1699){\makebox(0,0)[b]{{\SetFigFont{8}{9.6}{\rmdefault}{\mddefault}{\updefault}$\boldsymbol{S}_{D,1}[i]$}}}
\put(1662,1774){\ellipse{300}{300}}
\put(4662,1774){\ellipse{300}{300}}
\path(3012,1774)(3162,1774)
\blacken\path(3042.000,1744.000)(3162.000,1774.000)(3042.000,1804.000)(3042.000,1744.000)
\path(3837,1774)(3987,1774)
\blacken\path(3867.000,1744.000)(3987.000,1774.000)(3867.000,1804.000)(3867.000,1744.000)
\path(4137,1924)(4137,1624)
\path(3987,1774)(4287,1774)
\path(3012,499)(3162,499)
\blacken\path(3042.000,469.000)(3162.000,499.000)(3042.000,529.000)(3042.000,469.000)
\blacken\path(4167.000,1504.000)(4137.000,1624.000)(4107.000,1504.000)(4167.000,1504.000)
\path(4137,1624)(4137,1474)
\path(4287,1774)(4512,1774)
\blacken\path(4392.000,1744.000)(4512.000,1774.000)(4392.000,1804.000)(4392.000,1744.000)
\path(3162,1999)(3837,1999)(3837,1549)
	(3162,1549)(3162,1999)
\path(3162,724)(3837,724)(3837,274)
	(3162,274)(3162,724)
\path(2187,1999)(3012,1999)(3012,1549)
	(2187,1549)(2187,1999)
\path(2187,724)(3012,724)(3012,274)
	(2187,274)(2187,724)
\path(1812,1774)(2187,1774)
\blacken\path(2067.000,1744.000)(2187.000,1774.000)(2067.000,1804.000)(2067.000,1744.000)
\path(1287,1774)(1512,1774)
\blacken\path(1392.000,1744.000)(1512.000,1774.000)(1392.000,1804.000)(1392.000,1744.000)
\path(1512,1774)(1812,1774)
\path(1662,1924)(1662,1624)
\path(1662,2224)(1662,1924)
\blacken\path(1632.000,2044.000)(1662.000,1924.000)(1692.000,2044.000)(1632.000,2044.000)
\blacken\path(2067.000,469.000)(2187.000,499.000)(2067.000,529.000)(2067.000,469.000)
\path(2187,499)(1962,499)(1962,799)
\path(1962,1774)(1962,1474)
\blacken\path(1932.000,1594.000)(1962.000,1474.000)(1992.000,1594.000)(1932.000,1594.000)
\path(912,1999)(1287,1999)(1287,1549)
	(912,1549)(912,1999)
\path(537,1774)(912,1774)
\blacken\path(792.000,1744.000)(912.000,1774.000)(792.000,1804.000)(792.000,1744.000)
\path(237,1999)(537,1999)(537,1549)
	(237,1549)(237,1999)
\path(12,1774)(237,1774)
\blacken\path(117.000,1744.000)(237.000,1774.000)(117.000,1804.000)(117.000,1744.000)
\path(5487,1774)(5787,1774)
\blacken\path(5667.000,1744.000)(5787.000,1774.000)(5667.000,1804.000)(5667.000,1744.000)
\path(4962,1999)(5487,1999)(5487,1549)
	(4962,1549)(4962,1999)
\path(4962,1174)(5487,1174)(5487,724)
	(4962,724)(4962,1174)
\path(5637,1774)(5637,949)(5487,949)
\blacken\path(5607.000,979.000)(5487.000,949.000)(5607.000,919.000)(5607.000,979.000)
\path(4662,1924)(4662,1624)
\path(4962,949)(4662,949)(4662,1624)
\blacken\path(4692.000,1504.000)(4662.000,1624.000)(4632.000,1504.000)(4692.000,1504.000)
\path(4512,1774)(4962,1774)
\blacken\path(4842.000,1744.000)(4962.000,1774.000)(4842.000,1804.000)(4842.000,1744.000)
\path(3837,499)(4137,499)(4137,799)
\blacken\path(4167.000,679.000)(4137.000,799.000)(4107.000,679.000)(4167.000,679.000)
\put(2112,1999){\makebox(0,0)[b]{{\SetFigFont{8}{9.6}{\rmdefault}{\mddefault}{\updefault}$\boldsymbol{y}[i]$}}}
\put(4212,1324){\makebox(0,0)[b]{{\SetFigFont{8}{9.6}{\rmdefault}{\mddefault}{\updefault}$N_{T}\times 1$}}}
\put(1212,1099){\makebox(0,0)[b]{{\SetFigFont{8}{9.6}{\rmdefault}{\mddefault}{\updefault}$LN_{R}\times LN_{T}$}}}
\put(1887,1324){\makebox(0,0)[b]{{\SetFigFont{8}{9.6}{\rmdefault}{\mddefault}{\updefault}$LN_{R}\times 1$}}}
\put(3537,1324){\makebox(0,0)[b]{{\SetFigFont{8}{9.6}{\rmdefault}{\mddefault}{\updefault}$D\times 1$}}}
\put(2637,1174){\makebox(0,0)[b]{{\SetFigFont{8}{9.6}{\rmdefault}{\mddefault}{\updefault}$LN_{R}\times D$}}}
\put(1662,2224){\makebox(0,0)[b]{{\SetFigFont{8}{9.6}{\rmdefault}{\mddefault}{\updefault}$\boldsymbol{n}[i]$}}}
\put(3537,49){\makebox(0,0)[b]{{\SetFigFont{8}{9.6}{\rmdefault}{\mddefault}{\updefault}$D\times 1$}}}
\put(387,1699){\makebox(0,0)[b]{{\SetFigFont{8}{9.6}{\rmdefault}{\mddefault}{\updefault}Tx}}}
\put(687,1999){\makebox(0,0)[b]{{\SetFigFont{8}{9.6}{\rmdefault}{\mddefault}{\updefault}$\boldsymbol{x}_{T}[i]$}}}
\put(12,1999){\makebox(0,0)[b]{{\SetFigFont{8}{9.6}{\rmdefault}{\mddefault}{\updefault}$\boldsymbol{x}[i]$}}}
\put(687,2149){\makebox(0,0)[b]{{\SetFigFont{8}{9.6}{\rmdefault}{\mddefault}{\updefault}$LN_{T}\times 1$}}}
\put(162,1324){\makebox(0,0)[b]{{\SetFigFont{8}{9.6}{\rmdefault}{\mddefault}{\updefault}$N_{T}\times 1$}}}
\put(5187,1699){\makebox(0,0)[b]{{\SetFigFont{8}{9.6}{\rmdefault}{\mddefault}{\updefault}$Q(\cdot)$}}}
\put(5187,874){\makebox(0,0)[b]{{\SetFigFont{8}{9.6}{\rmdefault}{\mddefault}{\updefault}$\boldsymbol{F}[i]$}}}
\put(5187,499){\makebox(0,0)[b]{{\SetFigFont{8}{9.6}{\rmdefault}{\mddefault}{\updefault}$N_{T}\times N_{T}$}}}
\put(5712,1999){\makebox(0,0)[b]{{\SetFigFont{8}{9.6}{\rmdefault}{\mddefault}{\updefault}$\hat{\boldsymbol{x}}[i]$}}}
\put(1090,1699){\makebox(0,0)[b]{{\SetFigFont{8}{9.6}{\rmdefault}{\mddefault}{\updefault}$\boldsymbol{H}[i]$}}}
\put(3481,424){\makebox(0,0)[b]{{\SetFigFont{8}{9.6}{\rmdefault}{\mddefault}{\updefault}$\bar{\boldsymbol{w}}_{N_{T}}[i]$}}}
\put(4841,1990){\makebox(0,0)[b]{{\SetFigFont{8}{9.6}{\rmdefault}{\mddefault}{\updefault}$\boldsymbol{z}[i]$}}}
\put(4786,1431){\makebox(0,0)[b]{{\SetFigFont{8}{9.6}{\rmdefault}{\mddefault}{\updefault}$-$}}}
\put(4133,1029){\makebox(0,0)[b]{{\SetFigFont{8}{9.6}{\rmdefault}{\mddefault}{\updefault}$\vdots$}}}
\put(3527,993){\makebox(0,0)[b]{{\SetFigFont{8}{9.6}{\rmdefault}{\mddefault}{\updefault}$\vdots$}}}
\put(2602,958){\makebox(0,0)[b]{{\SetFigFont{8}{9.6}{\rmdefault}{\mddefault}{\updefault}$\vdots$}}}
\put(1943,888){\makebox(0,0)[b]{{\SetFigFont{8}{9.6}{\rmdefault}{\mddefault}{\updefault}$\vdots$}}}
\put(2588,431){\makebox(0,0)[b]{{\SetFigFont{8}{9.6}{\rmdefault}{\mddefault}{\updefault}$\boldsymbol{S}_{D,N_{T}}[i]$}}}
\put(3495,1699){\makebox(0,0)[b]{{\SetFigFont{8}{9.6}{\rmdefault}{\mddefault}{\updefault}$\bar{\boldsymbol{w}}_{1}[i]$}}}
\put(4145,1774){\ellipse{300}{300}}
\end{picture}
}